\documentclass[letterpaper,twocolumn,english,citesort,showpacs,showkeys,preprintnumbers,prd,floatfix,nofootinbib,superscriptaddress]{revtex4}
\usepackage[T1]{fontenc}
\usepackage[latin1]{inputenc}
\usepackage{graphicx}
\usepackage{amssymb}

\makeatletter

\providecommand{\tabularnewline}{\\}
\newcommand{\lyxdot}{.}

\@ifundefined{definecolor}
 {\usepackage{color}}{}
\@ifundefined{definecolor}
 {\usepackage{color}}{}
\usepackage{hyperref}

\makeatother

\makeatother

\usepackage{babel}
\makeatother

\usepackage{babel}
\makeatother

\begin{document}
\eprint{ LPSC 09-081; SMU-HEP 09-13 }

\title{
Parton Distribution Function
Nuclear Corrections for Charged Lepton 
\\
and 
Neutrino 
Deep Inelastic Scattering
 Processes}


\author{I.~Schienbein}

\thanks{schien@lpsc.in2p3.fr}

\affiliation{Laboratoire de Physique Subatomique et de Cosmologie, Université
Joseph Fourier/CNRS-IN2P3/INPG, \\
 53 Avenue des Martyrs, 38026 Grenoble, France}

\author{\ \ J.~Y.~Yu\,}

\thanks{yu@physics.smu.edu}

\affiliation{Southern Methodist University, Dallas, TX 75275, USA}

\author{\ K.~Kova\v{r}\'{\i}k}

\thanks{kovarik@lpsc.in2p3.fr}

\affiliation{Laboratoire de Physique Subatomique et de Cosmologie, Université
Joseph Fourier/CNRS-IN2P3/INPG, \\
 53 Avenue des Martyrs, 38026 Grenoble, France}

\author{\ C.~Keppel\,}

\thanks{keppel@jlab.org}

\affiliation{Thomas Jefferson National Accelerator Facility, Newport News, VA
23602, USA}

\affiliation{Hampton University, Hampton, VA, 23668, USA}

\author{\ J.~G.~Morf\'{\i}n\,}

\thanks{morfin@fnal.gov}

\affiliation{Fermilab, Batavia, IL 60510, USA}

\author{\ \ F.~I.~Olness \,}

\thanks{olness@smu.edu}

\affiliation{Southern Methodist University, Dallas, TX 75275, USA}

\author{\ \ J.F.~Owens\,}

\thanks{owens@hep.fsu.edu}

\affiliation{Florida State University, Tallahassee, FL 32306-4350, USA}



\begin{abstract}
We perform a $\chi^{2}$-analysis of Nuclear Parton Distribution Functions
(NPDFs) using neutral current charged-lepton $(\ell^{\pm}A)$ Deeply
Inelastic Scattering (DIS) and Drell-Yan data for several nuclear
targets. The nuclear $A$ dependence of the NPDFs is extracted in
a next-to-leading order fit. We compare the nuclear corrections factors
($F_{2}^{Fe}/F_{2}^{D}$) for this charged-lepton data with other
results from the literature. In particular, we compare and contrast
fits based upon the charged-lepton DIS data with those using neutrino-nucleon
DIS data. 
\end{abstract}
\maketitle
\tableofcontents{}

\section{Introduction\label{sec:intro}}

\subsection{PDFs and Nuclear Corrections}

Parton distribution functions (PDFs) are of supreme importance in
contemporary high energy physics as they are needed for the computation
of reactions involving hadrons based on QCD factorization theorems
\citep{Collins:1989gx,Collins:1987pm,Collins:1998rz}. For this reason
various groups present global analyses of PDFs for protons \citep{Ball:2009mk,Ball:2008by,Martin:2009iq,Martin:2007bv,Nadolsky:2008zw,Tung:2006tb,JimenezDelgado:2008hf,Gluck:2007ck}
and nuclei \citep{Hirai:2007sx,Hirai:2004wq,Eskola:2009uj,Eskola:2008ca,Eskola:2007my,deFlorian:2003qf}
which are regularly updated in order to meet the increasing demand
for precision. 
The PDFs are non-perturbative objects which must be determined by
experimental input. To fully constrain the $x$-dependence and flavor-dependence
of the PDFs requires large data sets from different processes which
typically include Deeply Inelastic Scattering (DIS), Drell--Yan (DY),
and jet production.

While some of this data is extracted from free protons, much is taken
from a variety of nuclear targets. Because the neutrino cross section
is so small, to obtain sufficient statistics for the neutrino-nuclear
DIS processes it is necessary to use massive targets (\emph{e.g.},
iron, lead, etc.). Therefore, nuclear corrections are required if
we are to include the heavy target data into the global analysis of
proton PDFs.

The heavy target neutrino DIS data plays an important role in extracting
the separate flavor components of the PDFs. In particular, this data
set gives the most precise information on the strange quark PDF. As
the strange quark uncertainty may limit the precision of particular
Large Hadron Collider (LHC) $W$ and $Z$ measurements, the nuclear
corrections and their uncertainties will have a broad impact on a
comprehensive understanding of current and future data sets.

\subsection{Nuclear Corrections in the Literature}

\begin{figure}
\includegraphics[width=0.45\textwidth]{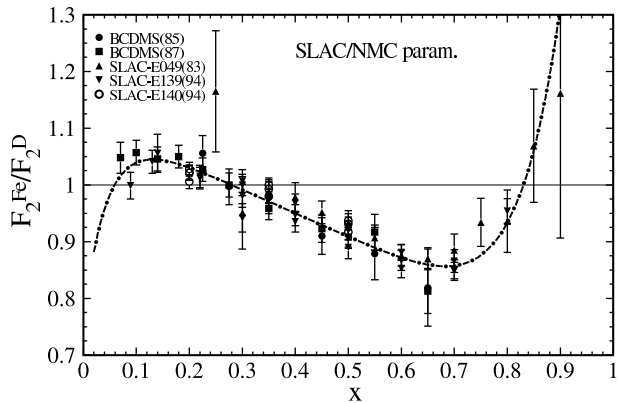}

\caption{Nuclear correction ratio, $F_{2}^{Fe}/F_{2}^{D}$, as a function of
$x$. The parameterized curve is compared to SLAC and BCDMS data \citep{Bodek:1983qn,Bari:1985ga,Benvenuti:1987az,Landgraf:1991nv,Gomez:1993ri,Dasu:1993vk,Rondio:1993mf}.
\label{fig:slac}}

\end{figure}

In previous PDF analyses \citep{Lai:1996mg,Lai:1999wy}, a fixed nuclear
correction was applied to {}``convert'' the data from a heavy target
to a proton. As such, these nuclear correction factors were frozen
at a fixed value. They did not adjust for the $Q^{2}$ scale or the
physical observable ($F_{2}$, $F_{3}$, $\frac{d\sigma}{dxdy}$),
and they did not enter the PDF uncertainty analysis.

While this approach may have been acceptable in the past given the
large uncertainties, improvements in both data and theory precision
demand comparable improvements in the treatment of the nuclear corrections.

Figure~\ref{fig:slac} displays the $F_{2}^{Fe}/F_{2}^{D}$ structure
function ratio as measured by the SLAC and BCDMS collaborations. The
SLAC/NMC curve is the result of an A-independent parametrization fit
to calcium and iron charged-lepton DIS data \citep{Bodek:1983qn,Bari:1985ga,Benvenuti:1987az,Landgraf:1991nv,Gomez:1993ri,Dasu:1993vk,Rondio:1993mf,Owens:2007kp}.
This parameterization was used to {}``convert'' heavy target data
to proton data, which then would be input into the global proton PDF
fit.%
\footnote{Technically, the heavy target data was scaled to a deuteron target,
and then isospin symmetry relations were used to obtain the corresponding
proton data. Deuteron corrections were used in certain cases. %
} The SLAC/NMC parmeterization was then applied to \emph{both} charged-lepton--nucleus
and neutrino--nucleus data, and this correction was taken to be independent
of the scale $Q$ and the specific observable $\{F_{2},F_{3},...\}$.
Recent work demonstrates that the parameterized approximation of Fig.~\ref{fig:slac}
is not sufficient and it is necessary to account for these details
\citep{Kulagin:2004ie,Kulagin:2007ju,Schienbein:2007fs}.

\subsection{Outline }

In this paper, we present a new framework for a global analysis of
nuclear PDFs (NPDFs) at Next-to-Leading-Order (NLO). An important
and appealing feature of this framework is that it naturally extends
the proton analysis by endowing the free fit parameters with a dependence
on the atomic number $A$. This will allow us to study proton and
nuclear PDFs simultaneously such that nuclear correction factors needed
for the proton analysis can be computed dynamically.

In Section~\ref{sec:framework}, we outline our method for the analysis,
specify the DIS and DY data sets, and present the $\chi^{2}$ of our
fit. In Section~\ref{sec:Nuclear-Corrections:}, we compute the nuclear
correction factors ($F_{2}^{Fe}/F_{2}^{D}$) for the fit to the $\ell^{\pm}A$
and DY data. In Section~\ref{sec:NC-and-CC}, we compare these results
to the nuclear correction factors ($F_{2}^{Fe}/F_{2}^{D}$) from the
$\nu A$ fit of Ref.~\citep{Schienbein:2007fs}. Finally, we summarize
our results in Section~\ref{sec:conclusion}.

\section{NPDF Global Analysis Framework\label{sec:framework}}

\begin{figure}
\includegraphics[width=0.45\textwidth]{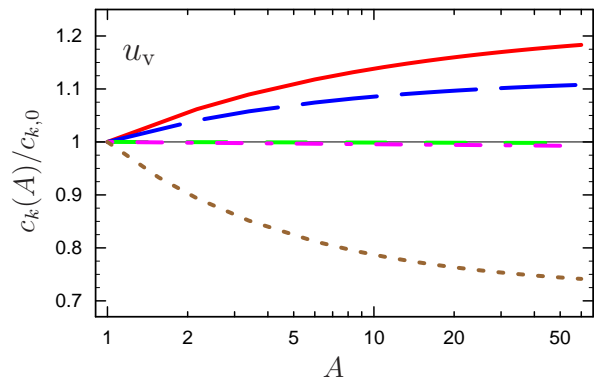} \\[5mm] \includegraphics[width=0.45\textwidth]{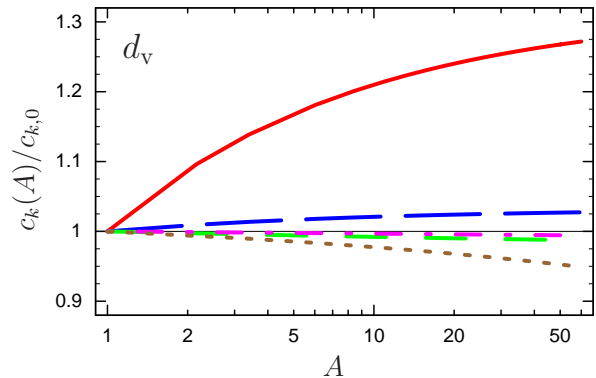}

\caption{We display the $A$-dependent coefficients $c_{k}(A)$, $k=\{1,5\}$,
for the up-valence (top) and down-valence PDF (bottom) as a function
of the nuclear $A$. The dependence of the coefficients $c_{k}(A)$
is shown by the following lines: $c_{1}$- solid (red) line, $c_{2}$-
long dashed (blue) line, $c_{3}$- dashed (green) line, $c_{4}$-
dash-dotted (magenta) line, $c_{5}$- dotted (brown) line.\label{fig:uValCoeffs}}

\end{figure}

\begin{figure*}
\includegraphics[width=0.45\textwidth]{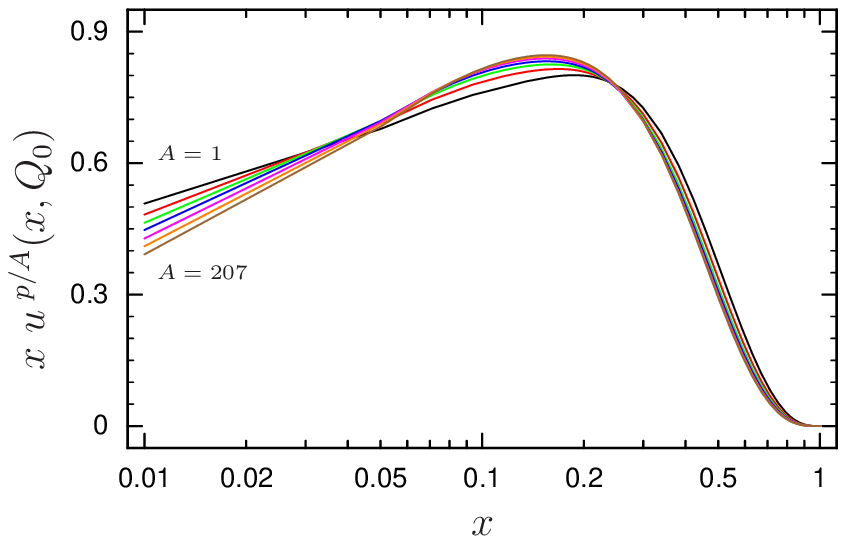}\hspace{0.05\textwidth}
\includegraphics[width=0.45\textwidth]{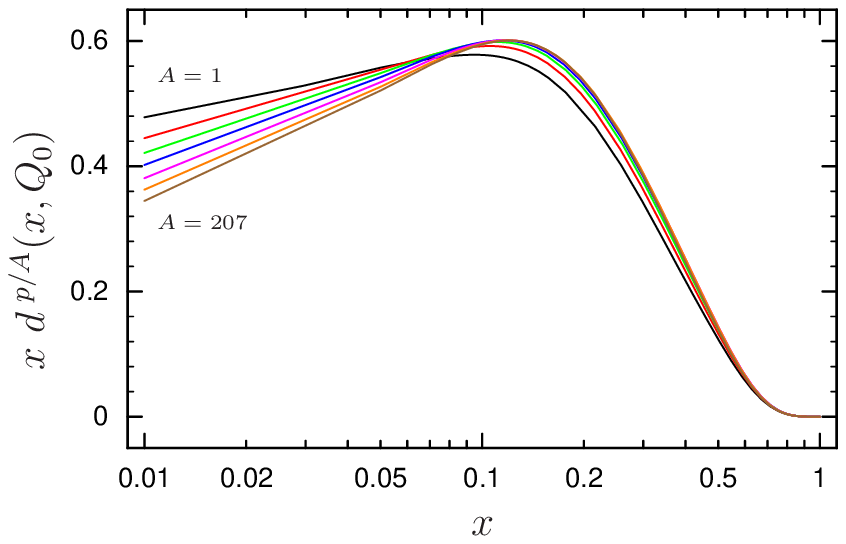}
\caption{We display the a) $x\,u(x)$ and b) $x\,d(x)$ PDFs for a selection of nuclear
$A$ values ranging from $A=\{1,207\}$. We choose $Q_{0}=1.3\,{\rm GeV}$.
The different curves depict the PDFs of nuclei with the following
atomic numbers (from top to bottom at $x=0.01$) $A=1,2,4,8,20,54,$
and $207$. \label{fig:udPDFs}}
\end{figure*}

\begin{table}[t]
 \begin{tabular}{|l|l|c|c|}
\hline 
{\scriptsize $\mathbf{F_{2}^{A}/F_{2}^{D}:}$ }  &  &  & \tabularnewline
{\scriptsize Observable}  & {\scriptsize Experiment }  & {\scriptsize Ref. }  & {\scriptsize $\#$ data }\tabularnewline
\hline
\hline 
{\scriptsize D }  & {\scriptsize NMC-97 }  & {\scriptsize \protect\citep{Arneodo:1996qe} }  & {\scriptsize 275 }\tabularnewline
{\scriptsize He/D }  & {\scriptsize SLAC-E139 }  & {\scriptsize \protect\citep{Gomez:1993ri} }  & {\scriptsize 18 }\tabularnewline
 & {\scriptsize NMC-95,re }  & {\scriptsize \protect\citep{Amaudruz:1995tq} }  & {\scriptsize 16 }\tabularnewline
 & {\scriptsize Hermes }  & {\scriptsize \protect\citep{Airapetian:2002fx}}  & {\scriptsize 92 }\tabularnewline
{\scriptsize Li/D }  & {\scriptsize NMC-95 }  & {\scriptsize \protect\citep{Arneodo:1995cs} }  & {\scriptsize 15 }\tabularnewline
{\scriptsize Be/D }  & {\scriptsize SLAC-E139 }  & {\scriptsize \protect\citep{Gomez:1993ri} }  & {\scriptsize 17 }\tabularnewline
{\scriptsize C/D }  & {\scriptsize EMC-88 }  & {\scriptsize \protect\citep{Ashman:1988bf} }  & {\scriptsize 9 }\tabularnewline
 & {\scriptsize EMC-90 }  & {\scriptsize \protect\citep{Arneodo:1989sy} }  & {\scriptsize 2 }\tabularnewline
 & {\scriptsize SLAC-E139 }  & {\scriptsize \protect\citep{Gomez:1993ri} }  & {\scriptsize 7 }\tabularnewline
 & {\scriptsize NMC-95,re }  & {\scriptsize \protect\citep{Amaudruz:1995tq} }  & {\scriptsize 16 }\tabularnewline
 & {\scriptsize NMC-95 }  & {\scriptsize \protect\citep{Arneodo:1995cs} }  & {\scriptsize 15 }\tabularnewline
 & {\scriptsize FNAL-E665-95 }  & {\scriptsize \protect\citep{Adams:1995is} }  & {\scriptsize 4 }\tabularnewline
{\scriptsize N/D }  & {\scriptsize BCDMS-85 }  & {\scriptsize \protect\citep{Bari:1985ga} }  & {\scriptsize 9 }\tabularnewline
 & {\scriptsize Hermes }  & {\scriptsize \protect\citep{Airapetian:2002fx}}  & {\scriptsize 92 }\tabularnewline
{\scriptsize Al/D }  & {\scriptsize SLAC-E049 }  & {\scriptsize \protect\citep{Bodek:1983ec} }  & {\scriptsize 18 }\tabularnewline
 & {\scriptsize SLAC-E139 }  & {\scriptsize \protect\citep{Gomez:1993ri} }  & {\scriptsize 17 }\tabularnewline
{\scriptsize Ca/D }  & {\scriptsize EMC-90 }  & {\scriptsize \protect\citep{Arneodo:1989sy} }  & {\scriptsize 2 }\tabularnewline
 & {\scriptsize SLAC-E139 }  & {\scriptsize \protect\citep{Gomez:1993ri} }  & {\scriptsize 7 }\tabularnewline
 & {\scriptsize NMC-95,re }  & {\scriptsize \protect\citep{Amaudruz:1995tq} }  & {\scriptsize 15 }\tabularnewline
 & {\scriptsize FNAL-E665-95 }  & {\scriptsize \protect\citep{Adams:1995is} }  & {\scriptsize 4 }\tabularnewline
{\scriptsize Fe/D }  & {\scriptsize BCDMS-85 }  & {\scriptsize \protect\citep{Bari:1985ga} }  & {\scriptsize 6 }\tabularnewline
 & {\scriptsize BCDMS-87 }  & {\scriptsize \protect\citep{Benvenuti:1987az}}  & {\scriptsize 10 }\tabularnewline
 & {\scriptsize SLAC-E049 }  & {\scriptsize \protect\citep{Bodek:1983qn} }  & {\scriptsize 14 }\tabularnewline
 & {\scriptsize SLAC-E139 }  & {\scriptsize \protect\citep{Gomez:1993ri} }  & {\scriptsize 23 }\tabularnewline
 & {\scriptsize SLAC-E140 }  & {\scriptsize \protect\citep{Dasu:1993vk} }  & {\scriptsize 6 }\tabularnewline
{\scriptsize Cu/D }  & {\scriptsize EMC-88 }  & {\scriptsize \protect\citep{Ashman:1988bf} }  & {\scriptsize 9 }\tabularnewline
 & {\scriptsize EMC-93(addendum)}  & {\scriptsize \protect\citep{Ashman:1992kv}}  & {\scriptsize 10 }\tabularnewline
 & {\scriptsize EMC-93(chariot)}  & {\scriptsize \protect\citep{Ashman:1992kv} }  & {\scriptsize 9 }\tabularnewline
{\scriptsize Kr/D }  & {\scriptsize Hermes }  & {\scriptsize \protect\citep{Airapetian:2002fx}}  & {\scriptsize 84 }\tabularnewline
{\scriptsize Ag/D }  & {\scriptsize SLAC-E139 }  & {\scriptsize \protect\citep{Gomez:1993ri} }  & {\scriptsize 7 }\tabularnewline
{\scriptsize Sn/D }  & {\scriptsize EMC-88 }  & {\scriptsize \protect\citep{Ashman:1988bf} }  & {\scriptsize 8 }\tabularnewline
{\scriptsize Xe/D }  & {\scriptsize FNAL-E665-92(em cut)}  & {\scriptsize \protect\citep{Adams:1992nf}}  & {\scriptsize 4 }\tabularnewline
{\scriptsize Au/D }  & {\scriptsize SLAC-E139 }  & {\scriptsize \protect\citep{Gomez:1993ri} }  & {\scriptsize 18 }\tabularnewline
{\scriptsize Pb/D }  & {\scriptsize FNAL-E665-95 }  & {\scriptsize \protect\citep{Adams:1995is} }  & {\scriptsize 4 }\tabularnewline
\hline
\hline 
\textbf{\scriptsize Total:}{\scriptsize {} }  &  &  & {\scriptsize 862 }\tabularnewline
\hline
\end{tabular}

\caption{{\scriptsize The DIS $F_{2}^{A}/F_{2}^{D}$ data sets used in the
fit. The table details the specific nuclear targets, references, and
the number of data points without kinematical cuts. }}

{\scriptsize \label{tab:exp1} } 
\end{table}

\begin{table}[t]
 \begin{tabular}{|l|l|c|c|}
\hline 
{\scriptsize $\mathbf{F_{2}^{A}/F_{2}^{A'}:}$ }  &  &  & \tabularnewline
{\scriptsize Observable }  & {\scriptsize Experiment }  & {\scriptsize Ref. }  & {\scriptsize $\#$ data }\tabularnewline
\hline
\hline 
{\scriptsize Be/C }  & {\scriptsize NMC-96 }  & {\scriptsize \protect\citep{Arneodo:1996rv} }  & {\scriptsize 15 }\tabularnewline
{\scriptsize Al/C }  & {\scriptsize NMC-96 }  & {\scriptsize \protect\citep{Arneodo:1996rv} }  & {\scriptsize 15 }\tabularnewline
{\scriptsize Ca/C }  & {\scriptsize NMC-95 }  & {\scriptsize \protect\citep{Amaudruz:1995tq} }  & {\scriptsize 20 }\tabularnewline
 & {\scriptsize NMC-96 }  & {\scriptsize \protect\citep{Arneodo:1996rv} }  & {\scriptsize 15 }\tabularnewline
{\scriptsize Fe/C }  & {\scriptsize NMC-95 }  & {\scriptsize \protect\citep{Arneodo:1996rv} }  & {\scriptsize 15 }\tabularnewline
{\scriptsize Sn/C }  & {\scriptsize NMC-96 }  & {\scriptsize \protect\citep{Arneodo:1996ru} }  & {\scriptsize 144 }\tabularnewline
{\scriptsize Pb/C }  & {\scriptsize NMC-96 }  & {\scriptsize \protect\citep{Arneodo:1996rv} }  & {\scriptsize 15 }\tabularnewline
{\scriptsize C/Li }  & {\scriptsize NMC-95 }  & {\scriptsize \protect\citep{Amaudruz:1995tq} }  & {\scriptsize 20 }\tabularnewline
{\scriptsize Ca/Li}  & {\scriptsize NMC-95 }  & {\scriptsize \protect\citep{Amaudruz:1995tq} }  & {\scriptsize 20 }\tabularnewline
\hline
\hline 
\textbf{\scriptsize Total:}{\scriptsize {} }  &  &  & {\scriptsize 279 }\tabularnewline
\hline
\end{tabular}

\caption{{\scriptsize The DIS $F_{2}^{A}/F_{2}^{A'}$ data sets used in the
fit. The table details the specific nuclear targets, references, and
the number of data points without kinematical cuts. }}

{\scriptsize \label{tab:exp2} } 
\end{table}

\begin{table}[t]
 \begin{tabular}{|l|l|c|c|}
\hline 
{\scriptsize $\mathbf{\sigma_{DY}^{pA}/\sigma_{DY}^{pA'}:}$ }  &  &  & \tabularnewline
{\scriptsize Observable }  & {\scriptsize Experiment }  & {\scriptsize Ref. }  & {\scriptsize $\#$ data }\tabularnewline
\hline
\hline 
{\scriptsize C/D }  & {\scriptsize FNAL-E772-90 }  & {\scriptsize \protect\citep{Alde:1990im} }  & {\scriptsize 9 }\tabularnewline
{\scriptsize Ca/D }  & {\scriptsize FNAL-E772-90 }  & {\scriptsize \protect\citep{Alde:1990im} }  & {\scriptsize 9 }\tabularnewline
{\scriptsize Fe/D }  & {\scriptsize FNAL-E772-90 }  & {\scriptsize \protect\citep{Alde:1990im} }  & {\scriptsize 9 }\tabularnewline
{\scriptsize W/D }  & {\scriptsize FNAL-E772-90 }  & {\scriptsize \protect\citep{Alde:1990im} }  & {\scriptsize 9 }\tabularnewline
{\scriptsize Fe/Be}  & {\scriptsize FNAL-E866-99 }  & {\scriptsize \protect\citep{Vasilev:1999fa} }  & {\scriptsize 28 }\tabularnewline
{\scriptsize W/Be }  & {\scriptsize FNAL-E866-99 }  & {\scriptsize \protect\citep{Vasilev:1999fa} }  & {\scriptsize 28 }\tabularnewline
\hline
\hline 
\textbf{\scriptsize Total:}{\scriptsize {} }  &  &  & {\scriptsize 92 }\tabularnewline
\hline
\end{tabular}

\caption{{\scriptsize The Drell-Yan data sets used in the fit. The table details
the specific nuclear targets, references, and the number of data points
without kinematical cuts. }}

{\scriptsize \label{tab:exp3} } 
\end{table}

\subsection{PDF analysis framework}

In this section, we present the global analysis of NPDFs using charged-lepton
DIS ($l^{\pm}A$) and Drell--Yan data to extend the analysis of Ref.~\citep{Owens:2007kp}
for a variety of nuclear targets. This analysis is performed in close
analogy with what is done for the $A=1$ free proton case \citep{Pumplin:2002vw}.
We will use the general features of the QCD-improved parton model
and the $\chi^{2}$ analyses as outlined in Ref.~\citep{Schienbein:2007fs}.
The input distributions are parameterized as \begin{widetext} \begin{eqnarray}
x\, f_{k}(x,Q_{0}) & = & c_{0}x^{c_{1}}(1-x)^{c_{2}}e^{c_{3}x}(1+e^{c_{4}}x)^{c_{5}}\qquad\quad k=u_{v},d_{v},g,\bar{u}+\bar{d},s,\bar{s}\,,\label{eq:input}\\
\bar{d}(x,Q_{0})/\bar{u}(x,Q_{0}) & = & c_{0}x^{c_{1}}(1-x)^{c_{2}}+(1+c_{3}x)(1-x)^{c_{4}}\,,\nonumber \end{eqnarray}
 \end{widetext} 
at the scale $Q_{0}=1.3$~GeV. Here, the $u_{v}$ and $d_{v}$ are
the up- and down-quark valence distributions, $\bar{u}$, $\bar{d}$,
$s$, $\bar{s}$ are the anti-up, anti-down, strange and anti-strange
sea distributions, and $g$ is the gluon.

In order to accommodate different nuclear target materials, we introduce
a nuclear $A$-dependence in the $c_{k}$ coefficients: \begin{equation}
c_{k}\to c_{k}(A)\equiv c_{k,0}+c_{k,1}\left(1-A^{-c_{k,2}}\right),\quad k=\{1,\ldots,5\}\,.\label{eq:Adep}\end{equation}
 This ansatz has the advantage that in the limit $A\to1$ we have
$c_{k}(A)\to c_{k,0}$; hence, $c_{k,0}$ is simply the corresponding
coefficient of the free proton. Thus, we can relate the $c_{k,0}$
parameters to the analogous quantities from proton PDF studies.

It is noteworthy that the $x$-dependence of our input distributions
$f_{k}^{p/A}(x,Q_{0})$ is the same for all nuclei $A$; hence, this
approach treats the NPDFs and the proton PDFs on the same footing.%
\footnote{The nuclear analogue of the scaling variable $x$ is defined as $x:=Ax_{A}$
where $x_{A}=Q^{2}/2P_{A}\cdot q$ is the usual Bjorken variable formed
out of the four-momenta of the nucleus ($P_{A}$) and the exchanged
boson ($q$), with $Q^{2}=-q^{2}$ \citep{Schienbein:2007fs}. %
} Additionally, this method facilitates the interpretation of the fit
at the parameter level by allowing us to study the $c_{k}(A)$ coefficients
as functions of the nuclear $A$ parameter.

With this $A$-generalized set of initial PDFs, we can apply the DGLAP
evolution equations to obtain the PDFs for a bound proton inside a
nucleus $A$, $f_{i}^{p/A}(x,Q)$ . We can then construct the PDFs
for a general $(A,Z)$-nucleus: \begin{equation}
f_{i}^{(A,Z)}(x,Q)=\frac{Z}{A}\ f_{i}^{p/A}(x,Q)+\frac{(A-Z)}{A}\ f_{i}^{n/A}(x,Q)\label{eq:pdf}\end{equation}
 where we relate the distributions of a bound neutron, $f_{i}^{n/A}(x,Q)$,
to those of a proton by isospin symmetry. Similarly, the nuclear structure
functions are given by: \begin{equation}
F_{i}^{(A,Z)}(x,Q)=\frac{Z}{A}\ F_{i}^{p/A}(x,Q)+\frac{(A-Z)}{A}\ F_{i}^{n/A}(x,Q)\,.\label{eq:sfs}\end{equation}
 These structure functions can be computed at next-to-leading order
as convolutions of the nuclear PDFs with the conventional Wilson coefficients,
\textit{i.e.}, generically \begin{equation}
F_{i}^{(A,Z)}(x,Q)=\sum_{k}C_{ik}\otimes f_{k}^{(A,Z)}\,.\label{eq:sfs2}\end{equation}
To account for heavy quark mass effects, we calculate the relevant
structure functions in the 
Aivazis-Collins-Olness-Tung
(ACOT) scheme \citep{Aivazis:1993kh,Aivazis:1993pi}
at NLO QCD \citep{Kretzer:1998ju}.

\subsection{Inputs to the Global NPDF Fit}

Using the above framework, we can then construct a global fit to charged-lepton--nucleus
($l^{\pm}A$) DIS data and Drell--Yan data. To guide our constraints
on the $c_{k,0}$ coefficients, we use the global fit of the proton
PDFs based upon Ref.~\citep{Owens:2007kp}. This fit has the advantage
that the extracted proton PDFs have minimal influence from nuclear
targets. To provide the $A$-dependent nuclear information, we use
a variety of $l^{\pm}A$ DIS data and Drell--Yan data. The complete
list of nuclear targets and processes is listed in Tables~\ref{tab:exp1},
\ref{tab:exp2}, and \ref{tab:exp3}; there are 1233 data points before
kinematical cuts are applied.

The structure of the fit is analogous to that of Ref.~\citep{Schienbein:2007fs}.
For the quark masses we take $m_{c}=1.3$~GeV and $m_{b}=4.5$~GeV.
To limit effects of higher-twist we choose standard kinematic cuts
of $Q_{cut}=2.0$~GeV, and $W_{cut}=3.5$~GeV as they are employed
in the CTEQ proton analyses.\footnote{For example, see the CTEQ 
(Coordinated Theoretical-Experimental project on QCD) 
analysis of Ref.~\citep{Pumplin:2002vw} which presents the CTEQ6 PDF sets.}
There are 708 data points which satisfy
these cuts. The fit was performed with 32 free parameters which gives
676 degrees of freedom (dof).

\subsection{Result of the NPDF Fit}

Performing the global fit to the data, we obtain an overall $\chi^{2}/{\rm dof}$
of 0.946. Individually, we find a $\chi^{2}/{\rm pt}$ of 0.919 for
the $F_{2}^{A}/F_{2}^{D}$ measurements of Table~\ref{tab:exp1},
of 0.685 for the $F_{2}^{A}/F_{2}^{A'}$ measurements of Table~\ref{tab:exp2},
and of 1.077 for the Drell--Yan measurements of Table~\ref{tab:exp3}.
The fact that we obtain a good fit implies that we have devised an
efficient parameterization of the underlying physics.

The output of the fit is the set of $c_{k,i}$ parameters and a set
of $A$-dependent momentum fractions for the gluon and the strange
quark. Using the $c_{k,i}$ coefficients we can construct the $A$-dependent
$c_{k}(A)$ functions which determine the nuclear PDFs at the initial
$Q_{0}$ scale: $f_{i}^{A}(x,Q_{0})$. As an example, we display the
$c_{k}(A)$ functions in Fig.~\ref{fig:uValCoeffs} for the case
of the up-valence and down-valence distributions.

Finally, we can use the DGLAP evolution equations to evolve to an
arbitrary $Q$ to obtain the desired $f_{i}^{A}(x,Q)$ functions.
In Fig.~\ref{fig:udPDFs} we display the up- and down-quark PDFs
at a scale of $Q_{0}=1.3$~GeV as a function of $x$ for a variety
of nuclear-$A$ values.


\section{\texorpdfstring{$\ell^{\pm}A$}{lA}
 Nuclear Corrections\label{sec:Nuclear-Corrections:}}

\begin{figure*}
\begin{picture}(500,155)(0,0) 
 \put(0,0){\includegraphics[width=0.45\textwidth]{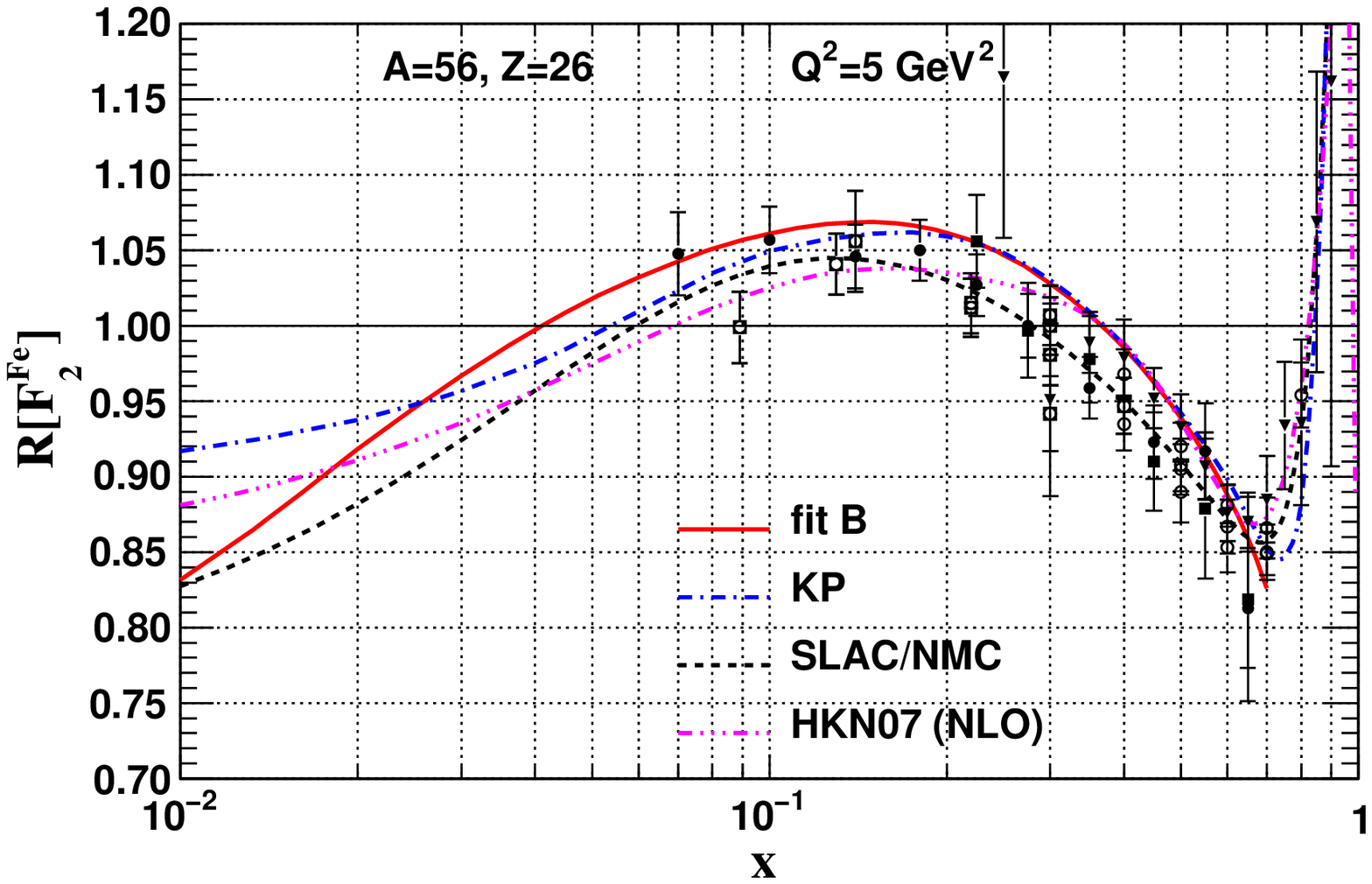}}
\put(260,0){\includegraphics[width=0.45\textwidth]{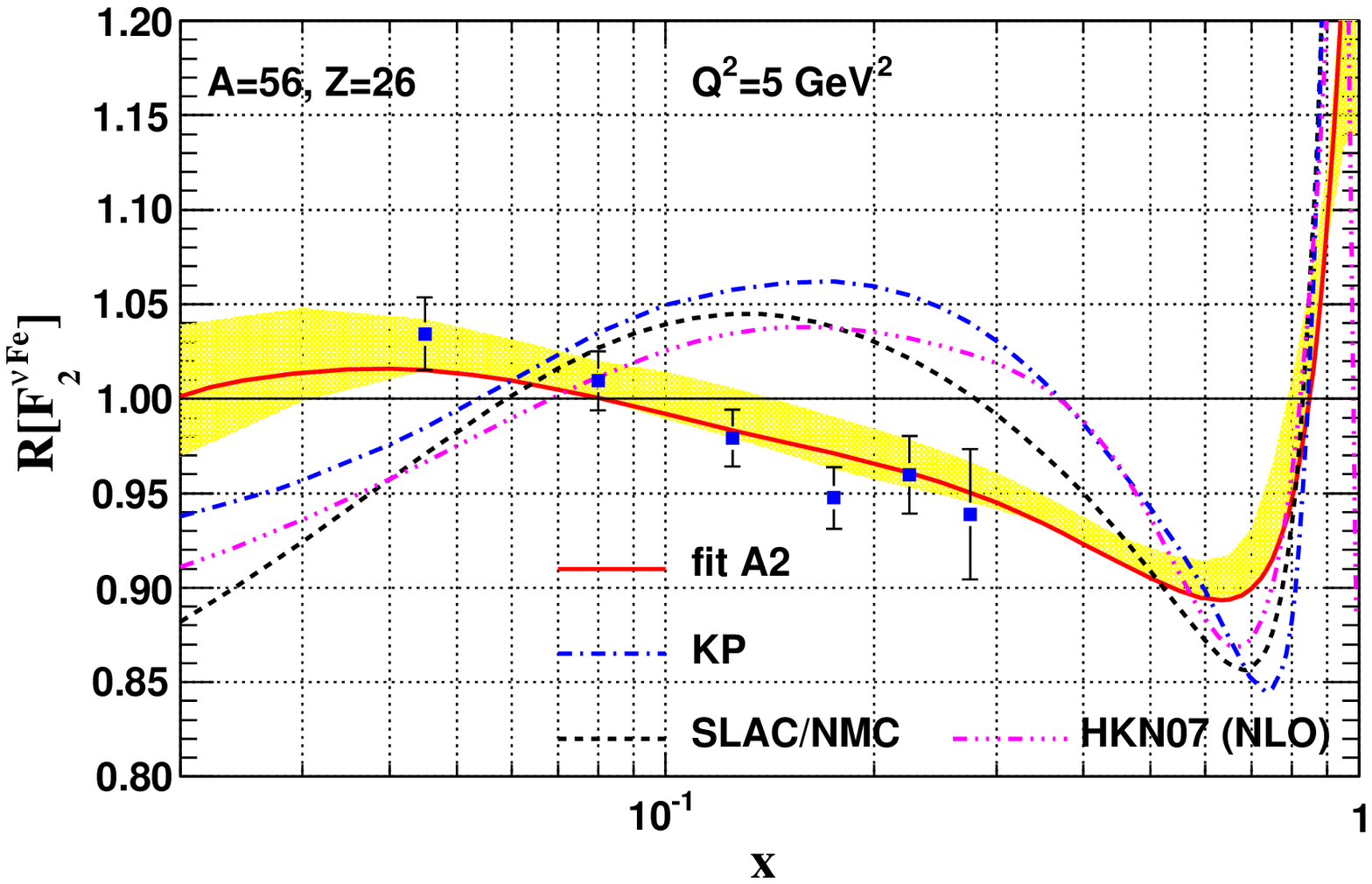}}
\put(114,0){$(a)$} \put(374,0){$(b)$} \end{picture} 

\caption{The computed nuclear correction ratio, $F_{2}^{Fe}/F_{2}^{D}$, as
a function of $x$ for $Q^{2}=5\,{\rm GeV}^{2}$. Figure-a) shows
the fit (fit B) using charged-lepton--nucleus ($\ell^{\pm}A$) and
DY data whereas Figure-b) shows the fit using neutrino-nucleus ($\nu A$)
data (fit A2 from Ref.~\citep{Schienbein:2007fs}). Both fits are
compared with the SLAC/NMC parameterization, as well as fits from
Kulagin-Petti (KP) (Ref.~\citep{Kulagin:2004ie,Kulagin:2007ju})
and Hirai et \textit{al.} (HKN07), (Ref.~\citep{Hirai:2007sx}).
The data points displayed in Figure-a) are the same as in Fig.~\ref{fig:slac}
and those displayed in Figure-b) come from the NuTeV experiment \citep{Tzanov:2005kr,TzanovPhD:200x}.
\label{fig:nuc05}}

\end{figure*}

\begin{figure*}
\begin{picture}(500,155)(0,0) 
 \put(0,0){\includegraphics[width=0.45\textwidth]{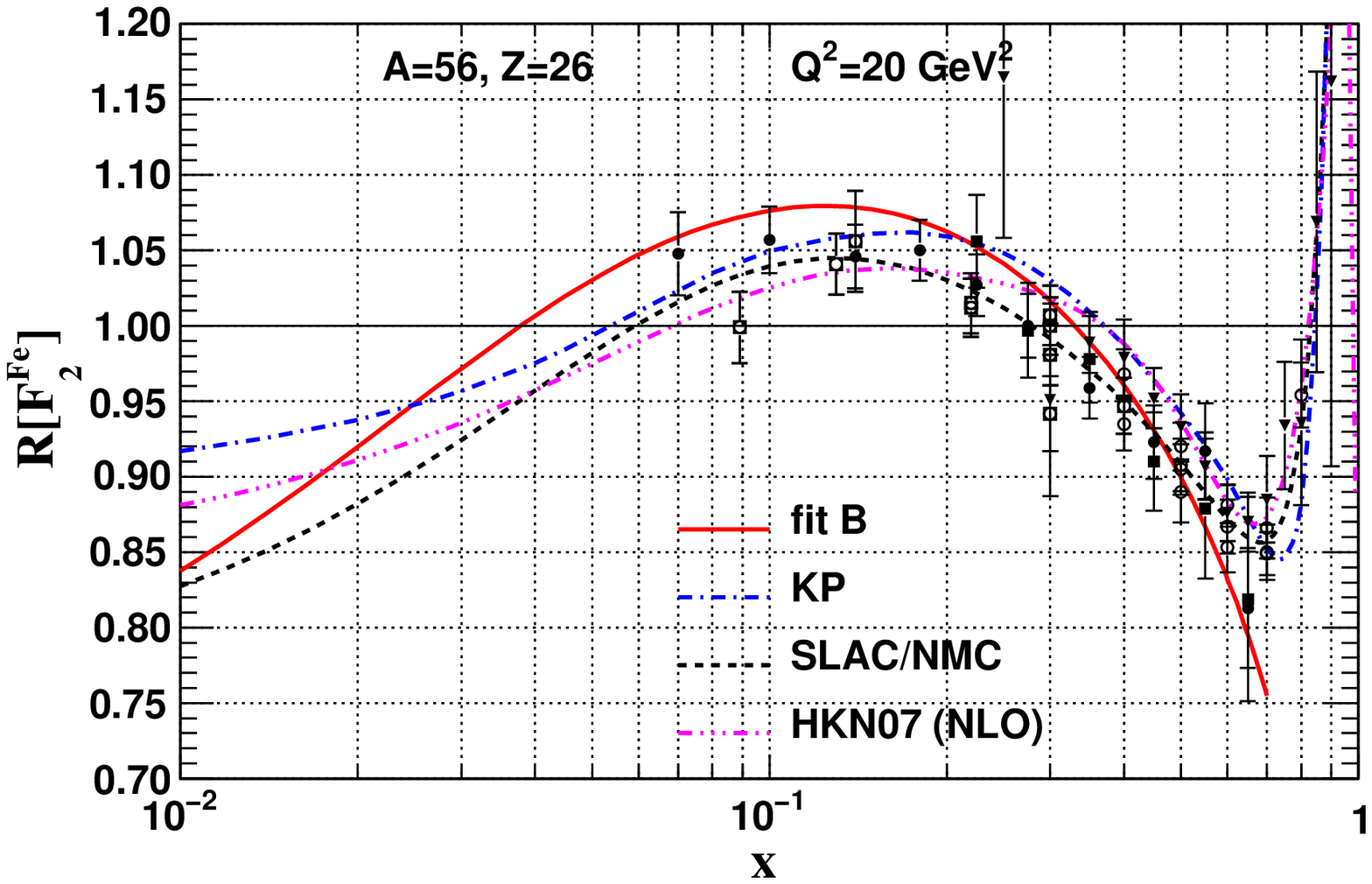}}
\put(260,0){\includegraphics[width=0.45\textwidth]{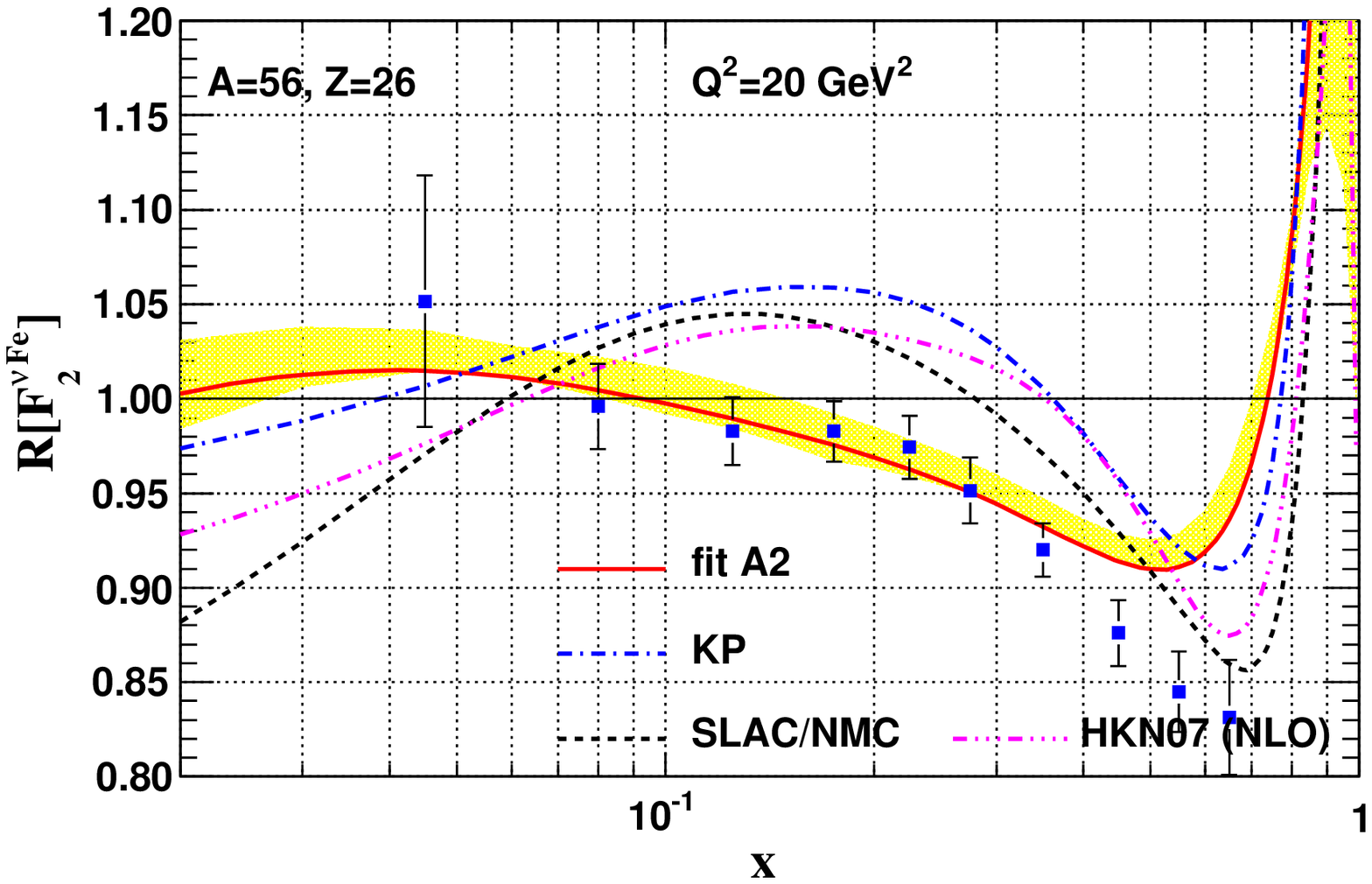}}
\put(114,0){$(a)$} \put(374,0){$(b)$} \end{picture}

\caption{Same as Figure~\ref{fig:nuc05} for $Q^{2}=20\,{\rm GeV}^{2}$.\label{fig:nuc20}}

\end{figure*}

Nuclear corrections are the key elements which allow us to combine
data across different nuclear targets and provide maximum information
on the proton PDFs. As the nuclear target data plays a critical role
in differentiating the separate partonic flavors (especially the strange
quark), this data provides the foundation that we will use to make
predictions at the LHC.

\subsection{Charged-Lepton (\texorpdfstring{$\ell^{\pm}A$}{lA}) Data}

The present nuclear PDF global analysis provides us with a complete
set of NPDFs $f_{i}^{A}(x,Q)$ with full functional dependence on
$\{x,Q,A\}$. Consequently, the traditional nuclear correction $F_{2}^{Fe}/F_{2}^{D}$
does not have to be applied as a {}``frozen'' external factor, but
can now become a dynamic part of the fit which can be adjusted to
accommodate the various data sets.

Having performed the fit outlined in Sec.~\ref{sec:framework}, we
can then use the $f_{i}^{A}(x,Q)$ to construct the corresponding
quantity $F_{2}^{Fe}/F_{2}^{D}$ to find the form that is preferred
by the data. In order to construct the ratio, we use the expression
given by Eq.~\ref{eq:sfs} for iron and deuterium. This result is
displayed in Figure~\ref{fig:nuc05}-a) for a scale of $Q^{2}=5$~GeV$^{2}$,
and in Figure~\ref{fig:nuc20}-a) for a scale of $Q^{2}=20$~GeV$^{2}$.
Comparing these figures, we immediately note that our ratio $F_{2}^{Fe}/F_{2}^{D}$
has non-trivial $Q$-dependence---as it should.

Figures~\ref{fig:nuc05}-a) and \ref{fig:nuc20}-a) also compare
our extracted $F_{2}^{Fe}/F_{2}^{D}$ ratio with the ($Q$-independent)
SLAC/NMC parameterization of Figure~\ref{fig:slac} and with the
fits from Kulagin-Petti (KP) \citep{Kulagin:2004ie,Kulagin:2007ju}
and Hirai-Kumano-Nagai (HKN07) \citep{Hirai:2007sx}. We observe that
in the intermediate range ($x\in\sim[0.07,0.7]$) where the bulk of
the SLAC/NMC data constrains the parameterization, our computed $F_{2}^{Fe}/F_{2}^{D}$
ratio compares favorably. When comparing the different curves, one
has to bear in mind the following two points. First, all curves in
principle have an uncertainty band which is not shown. Second, the
data points used to extract the SLAC/NMC curve are measured at different
$Q^{2}$ whereas our curve is always at a fixed $Q^{2}=5\,{\rm GeV}^{2}$
or $Q^{2}=20\,{\rm GeV}^{2}$. In light of these facts, we conclude
that our fit agrees very well with other models and parametrizations
as well as with the measured data points.

It should be noted that the kinematic cuts we employed to avoid higher
twist effects effectively exclude all data points in the high-$x$
region above $x\gtrsim0.7$. This is reflected by the fact that our
curves in Figs.~\ref{fig:nuc05}-a) and \ref{fig:nuc20}-a) stop
at $x=0.7$. The high-$x$ region is beyond the scope of this paper
and will be subject of a future analysis.

Thus, we find that data sets used in this fit ($F_{2}^{A}/F_{2}^{D}$,
$F_{2}^{A}/F_{2}^{A'}$, and $\sigma_{DY}^{pA}/\sigma_{DY}^{pA'}$
) are compatible with the SLAC, BCDMS, and NMC data. Additionally,
we can go further and use our complete set of NPDFs $f_{i}^{A}(x,Q)$
to compute the appropriate nuclear correction not only for $F_{2}^{Fe}/F_{2}^{D}$,
but for \emph{any} nuclear target ($A$) for any $Q$-value, and for
any observable. We make use of this property in the following section
where we compute the corresponding quantity for a different nuclear
process.

\section{\texorpdfstring{$\ell^{\pm}A$}{lA} and \texorpdfstring{$\nu A$}{vA} 
 Nuclear Corrections\label{sec:NC-and-CC}}

\subsection{Nuclear Corrections in \texorpdfstring{$\nu A$}{vA} DIS}

In a previous analysis \citep{Schienbein:2007fs}, we examined the
charged current (CC) neutrino--nucleus DIS process $\nu A\to\mu X$,
and extracted the $F_{2}^{Fe}/F_{2}^{D}$ ratio.%
\footnote{While Ref.~\protect\citep{Schienbein:2007fs} extracted the nuclear
PDFs using only the NuTeV neutrino--iron DIS data, Ref.~\citep{Owens:2007kp}
demonstrated that the Chorus neutrino--lead DIS data\citep{Onengut:2005kv}
was consistent with the NuTeV data set. %
}

These results are displayed in Figures~\ref{fig:nuc05}-b) and \ref{fig:nuc20}-b).
The solid line is the result of the global fit (fit A2), and this
is compared with the previous SLAC/NMC parameterization, as well as
fits KP and HKN07. The data points displayed come from the NuTeV experiment
\citep{Tzanov:2005kr,TzanovPhD:200x}. The (yellow) band is an approximation
of the uncertainty of the fits.

As observed above, the SLAC/NMC parameterization is generally consistent
with the results of KP and HKN as well as our B fit to $\ell^{\pm}A$
and DY data. However, the A2 fit of Figures~\ref{fig:nuc05}-b) and
\ref{fig:nuc20}-b) does not agree with any of these three results.
We now examine this in detail.

\subsection{\texorpdfstring{$\ell^{\pm}A$}{lA} and 
 \texorpdfstring{$\nu A$}{vA}  Comparison}

The contrast between the charged-lepton ($\ell^{\pm}A$) case and
the neutrino ($\nu A$) case is striking; while the charged-lepton
results generally align with the SLAC/NMC, KP, and HKN determinations,
the neutrino results clearly yield different behavior in the intermediate
$x$-region. We emphasize that both the charged-lepton and neutrino
results are not a model---they come directly from global fits to the
data. To emphasize this point, we have superimposed illustrative data
point in Figures~\ref{fig:nuc05}-b) and \ref{fig:nuc20}-b); these
are simply the $\nu A$ DIS data \citep{Tzanov:2005kr,TzanovPhD:200x}
scaled by the 
appropriate structure function, calculated with the
proton PDF of Ref.~\citep{Schienbein:2007fs}.

The mis-match between the results in charged-lepton and neutrino DIS
is particularly interesting given that there has been a long-standing
{}``tension'' between the light-target charged-lepton data and the
heavy-target neutrino data in the historical fits \citep{Botts:1992yi,Lai:1994bb}.
This study demonstrates that the tension is not only between charged-lepton
\emph{light-target} data and neutrino heavy-target data, but we now
observe this phenomenon in comparisons between neutrino and charged-lepton
\emph{heavy-target} data.

There are two possible interpretations of this result.

\begin{enumerate}
\item There is, in fact, a single {}``compromise'' solution for the $F_{2}^{Fe}/F_{2}^{D}$
nuclear correction factor which yields a good fit for both the $\nu A$
and $\ell^{\pm}A$ data. 
\item The nuclear corrections for the $\ell^{\pm}A$ and $\nu A$ processes
are different. 
\end{enumerate}
Considering possibility 1), the {}``apparent'' discrepancy observed
in Figures~\ref{fig:nuc05} and \ref{fig:nuc20} could simply reflect
uncertainties in the extracted nuclear PDFs. The global fit framework
introduced in this work paves the way for a unified analysis of the
$\ell^{\pm}A$, DY, and $\nu A$ data which will ultimately answer
this question. Having established the nuclear correction factors for
neutrino and charged-lepton processes separately, we can combine these
data sets (accounting for appropriate systematic and statistical errors)
to obtain a {}``compromise'' solution.%
\footnote{While it is straightforward to obtain a {}``fit'' to the combined
neutrino and charged-lepton DIS data sets, determining the appropriate
weights of the various sets and discerning whether this {}``compromise''
fit is within the allowable uncertainty range of the data is a more
involved task. This work is presently ongoing. %
}

If it can be established that a {}``compromise'' solution does not
exist, then the remaining option is that the nuclear corrections in
neutrino and charged-lepton DIS are different. This idea has previously
been discussed in the literature \citep{Brodsky:2004qa,Kulagin:2004ie,Kulagin:2007ju}.
We note that the charged-lepton processes occur (dominantly) via $\gamma$-exchange,
while the neutrino-nucleon processes occur via $W^{\pm}$-exchange.
Thus, the different nuclear corrections could simply be a consequence
of the differing propagation of the intermediate bosons (photon, $W$)
through dense nuclear matter. Regardless of whether this dilemma is
resolved via option 1) or 2), understanding this puzzle will provide
important insights about processes involving nuclear targets. Furthermore,
a deeper understanding could be obtained by a future high-statistics,
high-energy neutrino experiment using several nuclear target materials
\citep{Adams:2008cm,Adams:2009kp,Drakoulakos:2004qn}.

\section{Conclusions\label{sec:conclusion} }

We presented a new framework to carry out a global analysis of NPDFs
at next-to-leading order QCD, treating proton and nuclear targets
on equal footing. Within this approach, we have performed a $\chi^{2}$-analysis
of nuclear PDFs by extending the proton PDF fit of Ref.~\citep{Owens:2007kp}
to DIS $l^{\pm}A$ and Drell--Yan data. 
The result of the fit is a set of nuclear PDFs which incorporate
not only the $\{x,Q\}$-dependence,
but also the nuclear-$A$ degree of freedom; thus we can accommodate
the full range of nuclear targets from light ($A=1$) to heavy ($A=207$).
We find a good fit to the combined data set with a total $\chi^{2}/{\rm dof}$
of 0.946 demonstrating the viability of the framework.

We have used our results to compute the nuclear corrections factors,
and to compare these with the results from the literature. We find
good agreement for those fits based on a charged-lepton data set.

Separately, we have compared our nuclear corrections (derived with
a charged-lepton data set) with those computed using neutrino DIS
($\nu A\to\mu X$) data sets. Here, we observe substantive differences.

This fit is novel in several respects. 

\begin{itemize}
\item Since we constructed the nuclear PDF fits analogous to the proton
PDF fits, this framework allows a meaningful comparison between these
two distributions.

\item The above unified framework integrates the nuclear correction factors
as a dynamic component of the fit. These factors are essential if
we want to use the heavy target DIS data to constrain the strange
quark distribution of the proton, for example. 
\item This unified analysis of proton and nuclear PDFs provides the foundation
necessary to simultaneously analyze $\ell^{\pm}A$, DY and, $\nu A$
data. This will ultimately help in determining whether 1) a {}``compromise''
solution exists, or 2) the nuclear corrections depend on the exchanged
boson (e.g., $\gamma/Z$ or $W^{\pm}$). 
\end{itemize}
The compatibility of the charged-lepton $\ell^{\pm}A$ and neutrino-nucleus
$\nu A$ processes in the global analysis is an interesting and important
question. The resolution of this issue is essential for a complete
understanding of both the proton and nuclear PDFs. \vspace*{5mm}

\section*{Acknowledgment}

We thank 
Tim Bolton, 
Janet Conrad, 
Andrei Kataev,
Sergey Kulagin,
Shunzo Kumano, 
Dave Mason, 
W.~Melnitchouk,
Donna Naples, 
Roberto Petti,
Voica~A.~Radescu, 
Mary Hall Reno, 
and
Martin Tzanov 
for valuable discussions. F.I.O., I.S., and J.Y.Y.\ acknowledge
the hospitality of Argonne, BNL, CERN, Fermilab, and Les Houches where
a portion of this work was performed. This work was partially supported
by the U.S.\ Department of Energy under grant DE-FG02-04ER41299,
contract DE-FG02-97IR41022, contract DE-AC05-06OR23177 (under which
Jefferson Science Associates LLC operates the Thomas Jefferson National
Accelerator Facility), the National Science Foundation grant 0400332,
and the Lightner-Sams Foundation. The work of J.~Y.~Yu was supported
by the Deutsche Forschungsgemeinschaft (DFG) through grant No.~YU~118/1-1.
The work of K.~Kova\v{r}\'{\i}k was supported by the ANR projects
ANR-06-JCJC-0038-01 and ToolsDMColl, BLAN07-2-194882.

\bibliographystyle{hunsrt} \bibliographystyle{hunsrt}
\bibliography{npdf3}

\end{document}